\documentclass[
article,
twocolumn,
english,
footinbib,
tightenlines,
nobibnotes,
aps,
prb,
 unsortedaddress,
showpacs,
 superscriptaddress,
floatfix
]{revtex4-2}

\usepackage{natbib}
\usepackage[english]{babel}
\usepackage{letltxmacro}
\usepackage{latexsym}
\LetLtxMacro{\ORIGselectlanguage}{\selectlanguage}
\makeatletter
\DeclareRobustCommand{\selectlanguage}[1]{%
  \@ifundefined{alias@\string#1}
    {\ORIGselectlanguage{#1}}
    {\begingroup\edef\x{\endgroup
       \noexpand\ORIGselectlanguage{\@nameuse{alias@#1}}}\x}%
}
\newcommand{\definelanguagealias}[2]{%
  \@namedef{alias@#1}{#2}%
}
\makeatother
\definelanguagealias{en}{english}
\definelanguagealias{English}{english}
\usepackage{graphicx}
\usepackage{amsmath}
\usepackage{amsfonts}
\usepackage{amssymb}
\usepackage{bm}
\usepackage{color}
\usepackage[percent]{overpic}
\usepackage{soul} 
\usepackage{amssymb}
\usepackage{wasysym}
\usepackage{dsfont}
\usepackage{float}
\usepackage[mathscr]{euscript}
\usepackage{lipsum}
\usepackage{hyperref}
\hypersetup{
    pdfstartview={FitH},    
    colorlinks=true,       
    linkcolor=blue,          
    citecolor=blue,        
    filecolor=magenta,      
    urlcolor=blue           
}

\newcommand{\be}{\begin{equation}}
\newcommand{\ee}{\end{equation}}
\newcommand{\bea}{\begin{eqnarray}}
\newcommand{\eea}{\end{eqnarray}}

\usepackage{graphicx}
\usepackage[colorinlistoftodos]{todonotes}
\usepackage{verbatim}
\usepackage[normalem]{ulem}
\usepackage{enumitem}
\usepackage{multirow}

\usepackage{makerobust}

\begin{document}
\title{Engineering altermagnetic orders on the square-kagome lattice through sublattice interference}
\author{Jonas Issing}
\email{jonas.issing@uni-wuerzburg.de}
\affiliation{Institut f\"ur Theoretische Physik und Astrophysik and W\"urzburg-Dresden Cluster of Excellence ctd.qmat, Julius-Maximilians-Universit\"at W\"urzburg, Am Hubland, Campus S\"ud, W\"urzburg 97074, Germany}
\author{Jannis Seufert}
\affiliation{Institut f\"ur Theoretische Physik und Astrophysik and W\"urzburg-Dresden Cluster of Excellence ctd.qmat, Julius-Maximilians-Universit\"at W\"urzburg, Am Hubland, Campus S\"ud, W\"urzburg 97074, Germany}
\author{Michael Klett}
\affiliation{Institut f\"ur Theoretische Physik und Astrophysik and W\"urzburg-Dresden Cluster of Excellence ctd.qmat, Julius-Maximilians-Universit\"at W\"urzburg, Am Hubland, Campus S\"ud, W\"urzburg 97074, Germany}
\author{Sarbajit Mazumdar}
\affiliation{Institut f\"ur Theoretische Physik und Astrophysik and W\"urzburg-Dresden Cluster of Excellence ctd.qmat, Julius-Maximilians-Universit\"at W\"urzburg, Am Hubland, Campus S\"ud, W\"urzburg 97074, Germany}
\author{Yasir Iqbal}
\affiliation{Department of Physics and Quantum Centre of Excellence for Diamond and Emergent Materials (QuCenDiEM), Indian Institute of Technology Madras, Chennai 600036, India}
\author{Ronny Thomale}
\affiliation{Institut f\"ur Theoretische Physik und Astrophysik and W\"urzburg-Dresden Cluster of Excellence ctd.qmat, Julius-Maximilians-Universit\"at W\"urzburg, Am Hubland, Campus S\"ud, W\"urzburg 97074, Germany}
\author{Atanu Maity}
\email{atanu.maity@uni-wuerzburg.de}
\affiliation{Institut f\"ur Theoretische Physik und Astrophysik and W\"urzburg-Dresden Cluster of Excellence ctd.qmat, Julius-Maximilians-Universit\"at W\"urzburg, Am Hubland, Campus S\"ud, W\"urzburg 97074, Germany}
\begin{abstract}
 We investigate the emergence of altermagnetic (AM) phases on the square-kagome lattice. 
Our analysis reveals that matrix element effects due to an orthogonal sublattice weight decomposition of Fermi level eigenstates known as sublattice interference enable decoupled magnetic ordering tendencies on distinct sublattices. Depending on which sublattice undergoes a magnetic instability, we identify a $d_{xy}$-type AM phase and a $d_{x^{2}-y^{2}}$-type AM phase originating from different sublattice polarization patterns. Using the Kotliar-Ruckenstein slave boson formalism we explore the stability of these AM phases as a function of interaction strength. Our findings demonstrate that sublattice-selective magnetic instabilities provide a versatile route to engineer the nature of AM order.
   
\end{abstract}

\date{\today}
\maketitle
\section{Introduction}
Altermagnetism (AM) has recently emerged as a distinct class of collinear magnetic order, complementing ferromagnetism (FM) and antiferromagnetism (AFM). In FMs, magnetic moments align parallel, producing a net magnetization, whereas in AFMs neighboring spins align antiparallel and fully compensate. AM combines the traits of both the FM and AFM. It hosts an antiparallel spin arrangement but exhibits spin-polarized electronic structures without any net magnetization in real space. This coexistence of compensated magnetic moments and uncompensated momentum-space spin polarization is the defining character of AM and reflects a unique symmetry-breaking mechanism. From a symmetry point of view, AM arises when magnetic order breaks time-reversal symmetry in such a way that no anti-unitary symmetry combining time reversal with a spatial operation leaves crystal momentum invariant. In certain crystals, magnetic sublattices are related by point-group operations other than translations or inversions, preventing cancellation of spin polarization in momentum space. Consequently, although the macroscopic magnetization vanishes, the electronic bands develop a nontrivial spin texture with spin splitting and related transport effects. Space group-theoretical symmetry analysis and density functional theory (DFT) calculations~\cite{Libor-2022,Zeng-2024,Guo-2023,Roig-2024} establish this as the defining feature of the AM state. The identification of AM has renewed interest in compensated magnetic materials due to their potential for spintronics and quantum technologies. Unlike FMs, AM materials generate no stray fields, yet their spin-polarized bands can produce sizable spin currents and magneto-transport responses. This combination enables low-dissipation spintronic functionalities that merge compensated magnetic order with spin-polarized transport~\cite{Zarzuela-2025,Shi-2025,Guo-2025}.

Building on mean field treatments of crystal fields and/or magnetic order, theoretical predictions of AM first appeared in collinear magnetic metals with specific crystalline~\cite{Libor-2022,Zeng-2024} and later valley symmetries \cite{dürrnagel2026e}. Subsequent ab initio studies~\cite{Libor-2020,Krempasky-2024,Yang-2025,Guo-2023,Wan-2025} proposed candidate compounds such as MnTe, RuO$_2$, and CrSb, which display compensated antiferromagnetic configurations alongside pronounced spin splitting. Experimental evidence~\cite{Osumi-2024,Reimers-2024,Iguchi-2025,Feng-2022} from angle-resolved photoemission spectroscopy (ARPES), magneto-optical Kerr effect, and anomalous Hall measurements have partially confirmed AM behavior, nurturing the hope for rendering AM a robust and decently ubiquitous type of unconventional magnetic order. 

Despite this promising activity in the field, however, substantiating the microscopic origin of AM in correlated electron systems beyond mean field theory has proven to be surprisingly challenging. Early proposals attempted to  connect AM to orbital ordering~\cite{Leeb-2024,Wang-2025} or to structural distortions~\cite{Camerano-2025,Leon-2025}, coupling spin and orbital degrees of freedom. In an attempt to go beyond mean-field theory by proposing an orbital-order-driven mechanism for AM formation,~\cite{Leeb-2024} faced the crucial challenge of stabilizing staggered orbital order without inducing a net magnetization. This is because staggered two-orbital orders would tend to imply ferromagnetic spin exchange by the Goodenough Kanamori rules, and hence no staggered magnetic order. A multi-orbital ansatz might help to improve on this issue~\cite{kaushal2026spontaneousaltermagnetismmultiorbitalcorrelated}. Taking a different route, Ref.~\cite{Matteo-2025} proposed a microscopic model of itinerant electrons on a Lieb lattice that points to a qualitatively different microscopic mechanism: AM can arise from interacting electrons without any orbital or spin-orbit coupling, driven instead by sublattice interference~\cite{Kiesel-2012}--a purely kinetic effect encoded in the lattice electronic structure.  In that setting, the low-energy electronic states are strongly sublattice-polarized, so that the sublattice-resolved particle--hole excitations, and hence the magnetic fluctuation profile, attains a highly sublattice-dependent signature. Note that while this effect clearly has a strong bearing on sublattice-affected Fermi surface instabilities~\cite{Kiesel-2012,PhysRevLett.110.126405,PhysRevLett.127.177001,Wu-2023,Lin-2024,holbæk2026protectionunconventionalsuperconductivitydisorder}, {\it the sublattice interference is not constrained to the Fermi level and hence can also feature strong impact on phases descending from strongly correlated electron systems.} This enabled Ref.~\cite{Matteo-2025} to trigger a $\mathcal{PT}$-breaking AM instability for the Hubbard model on the Lieb lattice, stabilizing AM without invoking incipient conventional orbital or structural order.

Realizing that scattering channel selection through sublattice interference gives the most substantiated access to the spontaneous formation of AM in a correlated electron setting beyond mean field as shown for the Lieb lattice,  it readily suggests a broader paradigm in which AM can emerge generically in multi-sublattice metals with strong sublattice polarization. If different sublattices play distinct roles at low energies---for instance, one sublattice hosts the magnetic instability, while another supplies the additional symmetry environment needed for spontaneous $\mathcal{PT}$ breaking---then their interplay can produce an altermagnetic state with compensated net moment and momentum-dependent spin polarization.

Notably, sublattice polarization is quite generic in multi-sublattice metals \cite{Lin-2024}: at high-symmetry Brillouin zone points, point-group symmetry enforces destructive interference in Bloch states, suppressing wavefunction weight on select sublattices via cancellations of equivalent hopping paths. As filling is tuned, the Fermi level generically crosses these symmetry-organized bands, concentrating low-energy states on remaining sublattices—a robust kinetic effect persisting under moderate perturbations and longer-range hoppings.

Motivated by the attempt to elevate the sublattice fingerprint of correlated electron systems to a tuning knob for exotic electronic order, we investigate the realization of AM in the square–kagome lattice, a two-sublattice two-dimensional geometry with distinct coordination environments and electronic connectivity. 
The square–kagome lattice offers a suitable platform to demonstrate the spontaneous emergence of AM, as it contains two symmetry-inequivalent sublattices where AFM order on either one of these sublattices translates into AM order due to the appropriate symmetry relation between the sublattices~\cite{Che-2025}. The square–kagome lattice has previously been studied extensively in the localized-spin (Mott) regime — for instance in frustrated Heisenberg models and candidate quantum spin-liquid compounds such as $\mathrm{KCu_6AlBiO_4(SO_4)_5Cl}$~\cite{Fujihala-2020}, $\mathrm{KCu_7TeO_4(SO_4)_5Cl}$~\cite{gonzalez-2025}, ${\mathrm{Na}}_{6}{\mathrm{Cu}}_{7}{\mathrm{BiO}}_{4}{({\mathrm{PO}}_{4})}_{4}{\mathrm{Cl}}_{3}$~\cite{Niggemann-2023} etc. In contrast, its realization in an itinerant-electron setting remains largely unexplored. Here, we demonstrate that, in the itinerant regime, sublattice-dependent magnetic instabilities can generate multiple AM phases distinguished by their momentum-space spin polarization. Specifically, one sublattice predominantly drives a Néel-type instability, while the other facilitates the effective time-reversal symmetry breaking required for AM order to emerge.
We further employ the spin-rotationally invariant Kotliar-Ruckenstein (SRIKR) slave boson (SB)~\cite{{Kotliar-1986},{Fresard-1992}} formalism, which is particularly suited for multi-sublattice systems. The inclusion of Gaussian fluctuations enables a controlled assessment of the stability of the altermagnetic phases against charge and spin instabilities beyond static mean-field theory in the presence of strong interactions.

The paper is organized as follows. In Sec.~\ref{sec:symmetry_AM}, we classify the AM phases on the square–kagome lattice from symmetry standpoint and present corresponding minimal models. In Sec.~\ref{sec:model}, we introduce a tight-binding Hubbard model on this lattice. In Sec.~\ref{sec:spectrum}, we analyze the band structure of the tight-binding part of the model and discuss possible sublattice polarization scenarios. In Sec.~\ref{sec:am_orders}, we examine the AM instabilities that arise from sublattice polarization induced by the addition of on-site Hubbard terms. In Sec.~\ref{sec:stability_analysis}, we investigate the stability of the observed AM orders using the slave boson formalism. Finally, in Sec.~\ref{sec:discussion}, we summarize our results and provide an outlook.

\begin{figure}\includegraphics[width=1.0\linewidth]{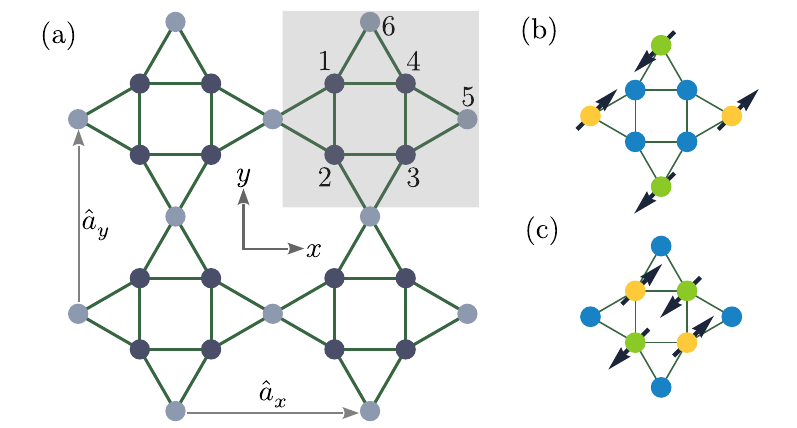}
\caption{(a) Illustration of the square-kagome lattice. The unit cell [marked by gray square box] consists of six sites indicated by numbers $1,2,\dots,6$. Among these, sites $1,2,3,4$ belong to Wyckoff position $4f$ while the sites $5,6$ correspond to the Wyckoff position $2c$. The unit translational vectors are denoted by $\hat{a}^{}_x$ and $\hat{a}^{}_y$. Possible AM phases due to the magnetization on (b) $2c$ sites and (c) $4f$ sites. Here, yellow and green sites denote two different magnetic alignments,  while blue sites denote nonmagnetic sites.}
\label{fig:fig1}
\end{figure}
\begin{figure}\includegraphics[width=1.0\linewidth]{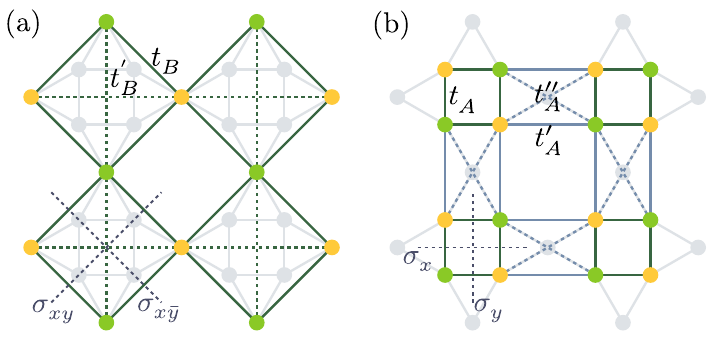}
\caption{Hopping amplitudes associated with the minimal models for the AM orders on (a) $2c$ and (b) $4f$ sites respectively.}
\label{fig:effective}
\end{figure}

\section{Symmetry-allowed AM phases}\label{sec:symmetry_AM}

\subsection{Lattice structure and symmetry classification}

The square-kagome lattice 
shown in Fig.~\ref{fig:fig1}(a),
belongs to the two-dimensional plane group $p4mm$ (no.~$11$), whose point group symmetry is $C_{4v}$. 
The lattice contains two inequivalent sublattices, denoted as A and B, occupying distinct Wyckoff positions within the unit cell. 
Sublattice A consists of sites 1, 2, 3, and 4 at the Wyckoff position $4f$ (general positions like $(x,x; x,\bar{x}; \bar{x},x;\bar{x},\bar{x})$ equivalents under $p4mm$), each with local site symmetry $m$. (mirror group ${1|m}$). These form the square corners of the lattice. 
Sublattice B consists of sites 5 and 6 at the Wyckoff position $2c$ (special positions like $(1/2,0; 0,1/2)$ equivalents under $p4mm$), each with local site symmetry $2mm$. (mirror group ${1, C_2| m, m}$). These occupy the kagome centers of the lattice.
The inequivalence of these Wyckoff positions leads to distinct local coordination environments, which play a central role in stabilizing different sublattice-polarized electronic states giving rise to AM phases. 

Within the point-group symmetry $C_{4v}$, AM order parameters are classified according to their transformation properties. In particular, the $d_{x^{2}-y^{2}}$-type AM order transforms as the $B_{1}$ irreducible representation, while the $d_{xy}$-type AM order belongs to the $B_{2}$ representation. These two symmetry channels correspond to distinct real-space patterns of sublattice magnetization and lead to different momentum-dependent spin polarizations in the electronic structure.

In the square-kagome lattice, sublattice polarization selects the realized symmetry channel. 
When magnetic instability 
develops on sublattice B (Wyckoff $2c$), while sublattice A remains non-magnetic, as shown in Fig.~\ref{fig:fig1}(b), the resulting AM phase transforms as $B_{1}$, corresponding to a $d_{x^{2}-y^{2}}$-type order. 
Conversely, when instability
develops on sublattice A (Wyckoff $4f$) as shown in Fig.~\ref{fig:fig1}(c), the system favors a $d_{xy}$-type AM phase associated with the $B_{2}$ representation.

\subsection{Spin group characterization and minimal models}
Collinear magnetic phases are characterized by a nontrivial spin group~\cite{Zeng-2024} denoted by $R_s=[R_i||R_j]$, where $R_i(R_j)$ incorporates all the spin (crystallographic) space operations. 
For collinear AM phases the nontrivial spin group is given by $R^{AM}_s=[E||H]+[C_2||G-H=\mathcal{A}]$ where $H$ is a halving subgroup of the layer group $G$. 
We analyze the non-relativistic spin-split band structures of these AM phases using spin-group formalism.
\subsubsection{AM sublattice order on $2c$ Wyckoff Sites}
The nontrivial spin group in this AM phase is given by $[C_2||\mathcal{A}]$ where $\mathcal{A}$ includes two diagonal mirrors $\sigma^{}_{{xy}}$ (mirror line $x=y$) and $\sigma^{}_{{x\bar{y}}}$ (mirror line $x=-y$), and a $C^{}_{4z}$ axis ($C^{}_4$) anchored at the center of the squares. This symmetry setting reshapes the spin-dependent band structure $\varepsilon^{}_{\sigma}(k^{}_x,k^{}_y)$ as follows.
\begin{equation}
\begin{aligned}
        [C^{}_2||\sigma^{}_{{x{y}}}]:\;&\varepsilon^{}_{\sigma}(k^{}_x,k^{}_y)\rightarrow\varepsilon^{}_{-\sigma}(k^{}_y,k^{}_x) \,, \\
        [C^{}_2||\sigma^{}_{{x\bar{y}}}]:\;&\varepsilon^{}_{\sigma}(k^{}_x,k^{}_y)\rightarrow\varepsilon^{}_{-\sigma}(-k^{}_y,-k^{}_x) \,, \\
        [C^{}_2||C^{}_{4}]:\;&\varepsilon^{}_{\sigma}(k^{}_x,k^{}_y)\rightarrow\varepsilon^{}_{-\sigma}(-k^{}_y,k^{}_x) \,.
\end{aligned}
\end{equation}
The first two relations enforce symmetry-protected spin degeneracy along the lines $k^{}_{x}=\pm k^{}_{y}$, i.e., along the segment $\Gamma$-$M$ in momentum space. 
The minimal model leading to such an AM phase can be cast as:
\begin{equation}
\label{eq:eff_BB}
H^B_{\text{eff}}=\varepsilon^{}_{0}(\mathbf{k})+d^{}_{x}(\mathbf{k})\tau^x+d^{}_{z}(\mathbf{k})\tau^z+M^{}_B\sigma^z\tau^z\,
\end{equation}
where $\tilde{a}=a+b$ with $a$ ($b$) as intra (inter) square length in Fig.~\ref{fig:fig1}(a) and 
\begin{equation}
\left.\begin{aligned}
&\varepsilon^{}_{0}(\mathbf{k})=\tilde{\mu}^{}_{B}-t^{\prime}_B+t^{\prime}_B(\cos^2\frac{k_x\tilde{a}}{2}+\cos^2\frac{k_y\tilde{a}}{2})\\
&d^{}_{z}(\mathbf{k})=t^{\prime}_B(\cos^2\frac{k_x\tilde{a}}{2}-\cos^2\frac{k_y\tilde{a}}{2})\\
&d^{}_{x}(\mathbf{k})=t^{}_B\cos\frac{k_x\tilde{a}}{2}\cos\frac{k_y\tilde{a}}{2}\,.
\end{aligned}\right.
\end{equation}
The Pauli matrices $\tau^\gamma$ ($\gamma=x,y,z$) label the two oppositely aligned magnetic sublattices, while $\sigma^\gamma$ denote spin degrees of freedom. The hoppings in this minimal model has been shown in Fig.~\ref{fig:effective}(a). Using a low-energy expansion, the momentum-dependent spin splitting around $\Gamma (0,0)$ point reads
\begin{equation}
    \label{eq:spinspliting_BB}
    f^{}_\Gamma(\mathbf{k})\propto (k^2_x-k^2_y)\,.
\end{equation}

It is worth noting that when this AM order arising from magnetization at the $2c$ Wyckoff position with site symmetry $2mm$ is embedded in three dimensions, it corresponds to an AM order at the $2e$ or $2f$ Wyckoff positions with site symmetry $D_{2h}$ of space group No.~123, according to the classification table of AM orders in Ref.~\cite{Roig-2024}.
\subsubsection{AM sublattice order on $4f$ Wyckoff Sites}
The crystallographic operations $\mathcal{A}$ connecting $\uparrow$ and $\downarrow$ sites include the horizontal ($\sigma_x$) and vertical ($\sigma_y$) mirrors and the rotational axis $C^{}_4$ ($C^{}_{4z}$) and yield the following constraints on the dispersion.
\begin{equation}
\begin{aligned}
        [C^{}_2||\sigma^{}_{x}]:\;&\varepsilon^{}_{\sigma}(k^{}_x,k^{}_y)\rightarrow\varepsilon^{}_{-\sigma}(k^{}_x,-k^{}_y) \,, \\
        [C^{}_2||\sigma^{}_{y}]:\;&\varepsilon^{}_{\sigma}(k^{}_x,k^{}_y)\rightarrow\varepsilon^{}_{-\sigma}(-k^{}_x,k^{}_y) \,, \\
        [C^{}_2||C^{}_{4}]:\;&\varepsilon^{}_{\sigma}(k^{}_x,k^{}_y)\rightarrow\varepsilon^{}_{-\sigma}(-k^{}_y,k^{}_x) \,.
\end{aligned}
\end{equation}
The first two relations ensure symmetry-protected spin degeneracy along the lines $k^{}_{x}=0,\pi$ and $k^{}_{y}=0,\pi$. 
In this case, four sites per unit cell form two pairs of oppositely spin-polarized sites, requiring an additional set of Pauli matrices $\Sigma^\gamma$ to distinguish sites with identical polarization. 
The minimal model reads
\begin{equation}
\label{eq:eff_AA}
H^A_{\text{eff}}=\varepsilon^{}_{0}(\mathbf{k})+d^{}_{\alpha\beta}(\mathbf{k})\tau^\alpha\Sigma^\beta+M^{}_A\sigma^z\tau^z\,
\end{equation}
where, $\varepsilon^{}_{0}(\mathbf{k})=\tilde{\mu}_A$ and the non-vanishing $d^{}_{\alpha\beta}(\mathbf{k})$ elements are given by:
\begin{equation}
\left.\begin{aligned}
&d^{}_{0x}=t^{\prime\prime}_A(\cos{k^{}_{x}b}\cos{k^{}_{y}a}+\cos{k^{}_{y}b}\cos{k^{}_{x}a})\\
&d^{}_{zx}=-t^{\prime\prime}_A(\sin{k^{}_{x}b}\sin{k^{}_{y}a}+\sin{k^{}_{y}b}\sin{k^{}_{x}a})\\
&d^{}_{0y}=t^{\prime\prime}_A(\sin{k^{}_{x}b}\cos{k^{}_{y}a}-\cos{k^{}_{y}b}\sin{k^{}_{x}a})\\
&d^{}_{zy}=t^{\prime\prime}_A(\cos{k^{}_{x}b}\sin{k^{}_{y}a}-\sin{k^{}_{y}b}\cos{k^{}_{x}a})\\
&d^{}_{x0}=t^{}_A\cos{k^{}_{y}a}+t^{\prime}_A\cos{k^{}_{y}b}\\
&d^{}_{yz}=t^{}_A\sin{k^{}_{y}a}-t^{\prime}_A\sin{k^{}_{y}b}\\
&d^{}_{xx}=t^{}_A\cos{k^{}_{x}a}+t^{\prime}_A\cos{k^{}_{x}b}\\
&d^{}_{yx}=-t^{}_A\sin{k^{}_{x}a}+t^{\prime}_A\sin{k^{}_{x}b}\,.\\
\end{aligned}\right.
\end{equation}
The hoppings in this minimal model
are
depicted in Fig.~\ref{fig:effective}(b). 
The long-wavelength expansion yields
\begin{equation}
    \label{eq:spinspliting_AA}
    f^{}_\Gamma(\mathbf{k})\propto k^{}_xk^{}_y\,.
\end{equation}
 Unlike the $B_1$ case (2c, minimal, mult. 2), this $B_2$ order embeds in SG 123 at 4j/4k (mult. 4), beyond standard prototypes~\cite{Roig-2024}. 

\section{Model}
\label{sec:model}
\begin{figure*}\includegraphics[width=1.0\linewidth]{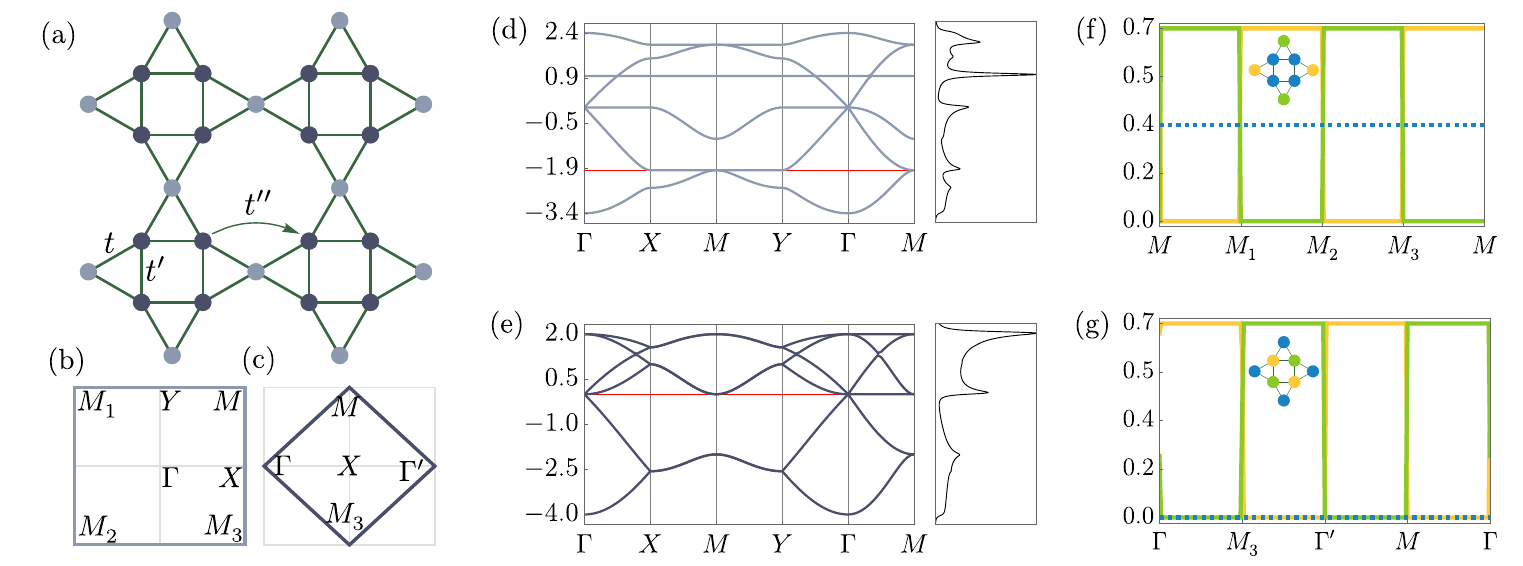}
	\caption{(a) Illustration of the tight-binding hopping parameters: $t$ (A-B hopping), $t^\prime$ (intra-unit-cell A-A hopping), and $t^{\prime\prime}$ (inter-unit-cell A-A hopping). (b) and (c) show the Fermi surfaces for $(t,t^\prime,t^{\prime\prime})=(1,0.5,0)$ and $(t,t^\prime,t^{\prime\prime})=(1,0.5,0.5)$ at fillings $n^{}_{f}\approx0.35$ and $n^{}_{f}\approx0.68$, respectively. The corresponding band dispersions are presented in (d) and (e), where the red line indicates the Fermi level (filling). The density of states (DOS) associated with each dispersion is shown on the right. For these two fillings, (f) and (g) display the sublattice weights along the paths $M\rightarrow M_1\rightarrow M_2\rightarrow M_3\rightarrow M$ and $\Gamma\rightarrow M_3\rightarrow\Gamma^\prime\rightarrow M\rightarrow\Gamma$, respectively.}
	\label{fig:fig2}
\end{figure*}
In order to realize the nontrivial spin–group characterized AM phases discussed in the previous section, it is essential to identify a minimal microscopic model that captures the underlying sublattice-polarization mechanism. 
To this end, we consider a single-orbital Hubbard model on the square–kagome lattice,
\begin{align}
&  \hat{H} = \hat{H}_0+\hat{H}_{\text{int}}\label{eq:model},\\
& \hat{H}_0 = - \sum_{\langle i,j \rangle_{ss'}, \sigma}
  t^{ss'}_{ij}  \left( \hat{c}_{i,s,\sigma}^\dagger \hat{c}_{j,s',\sigma} + \text{h.c.} \right)
    + \sum_{i,s,\sigma} \mu^{}_{i,s}\hat{n}_{i,s,\sigma}\label{eq:model_0},\\
& \hat{H}_{\text{int}} = U \sum_{i,s} \hat{n}_{i,s,\uparrow} \hat{n}_{i,s,\downarrow},\label{eq:model_int}
\end{align}
where $\hat{c}_{i,s,\sigma}^\dagger$ ($\hat{c}_{i,s,\sigma}$) creates (annihilates) an electron with spin $\sigma$ at the sublattice $s$ within the unit cell $i$, and $\hat{n}_{i,s,\sigma} = \hat{c}_{i,s,\sigma}^\dagger \hat{c}_{i,s,\sigma}$.
$\hat{H}_0$ represents the non-interacting part where the first term describes electron hopping on the square–kagome network with three distinct hopping amplitudes $t$, $t'$, and $t''$ as illustrated in Fig.~\ref{fig:fig2}(a) where we set $t=1$ in the rest of our discussion. By introducing the second term in $\hat{H}_0$, we include an intrinsic detuning (i.e., $\mu^{}_{i,s\in A}=\mu^{}_{ A}\neq\mu^{}_{i,s\in B}=\mu^{}_{ B}$) to incorporate the effect of different atomic species at Wyckoff positions $4f$ (A sites) and $2c$ (B sites).
The remaining term denoted by $H_{\text{int}}$ represents the on-site Hubbard interaction $U$.

\section{Electronic Spectrum of the Parent Metal and Sublattice Polarization}
\label{sec:spectrum}
In this section, we analyze the electronic band structure of the noninteracting (parent) square–kagome metal and highlight the nontrivial behavior of its Bloch eigenstates, which arises despite the full space-group symmetry of the single-particle Hamiltonian. We first focus on the case with $t^{\prime\prime}=0$. The tight-binding spectrum is obtained by Fourier transforming Eq.~\eqref{eq:model_0} and diagonalizing the resulting Bloch Hamiltonian, yielding
\begin{equation}
    \hat{H}_0 = \sum_{\mathbf{k},n,\sigma} 
    \varepsilon^{}_{\mathbf{k},n}\, 
    \hat{\gamma}^{\dagger}_{\mathbf{k},n,\sigma}
    \hat{\gamma}^{}_{\mathbf{k},n,\sigma},
\end{equation}
where $n$ labels the electronic bands and $\hat{\gamma}^{}_{\mathbf{k},n,\sigma}$ denotes the fermionic eigenmode. The latter is related to the sublattice operator $\hat{c}^{}_{\mathbf{k},s,\sigma}$ through the unitary transformation
\begin{equation}
    \hat{c}^{}_{\mathbf{k},s,\sigma} 
    = \sum_{n} \mathcal{U}^{s,n}_{\mathbf{k}}\,
    \hat{\gamma}^{}_{\mathbf{k},n,\sigma},
\end{equation}
where $\mathcal{U}^{s,n}_{\mathbf{k}}$ diagonalizes Eq.~\eqref{eq:model_0}. Figure~\ref{fig:fig2}(d) displays the resulting band structure for $t^{\prime}=0.5$, along with the corresponding density of states (DOS). Multiple van Hove singularities (VHSs) are present; the second-lowest VHS, whose Fermi level is indicated by the red line, exhibits particularly interesting features in the Bloch eigenmodes. To analyze this behavior, we express the real-space operators as
\begin{equation}
    \hat{c}^{}_{i,s,\sigma} 
    = \sum_{\mathbf{k},n}
      \mathcal{U}^{s,n}_{\mathbf{k}}\,
      \hat{\gamma}^{}_{\mathbf{k},n,\sigma}\,
      e^{i\mathbf{k}\cdot\mathbf{R}_{i,s}},
\end{equation}
where $\mathbf{R}_{i,s}$ denotes the position of site $(i,s)$. The essential information governing the nontrivial character of the Bloch states—and thereby the interacting instabilities—is encoded in the transformation coefficients $\mathcal{U}^{s,n}_{\mathbf{k}}$, which is referred to as the \textit{sublattice weights} following Refs.~\cite{Kiesel-2012,Wu-2023,Schwemmer-2024}.

At filling $n_{f}\approx0.35$ (marked by the red line in Fig.~\ref{fig:fig2}(d)), the Fermi level intersects a flat band along the Brillouin zone (BZ) boundary, giving rise to a Fermi surface (FS) shown in Fig.~\ref{fig:fig2}(b). The sublattice-resolved spectral weights $|\mathcal{U}^{s,n}_{\mathbf{k}}|$ are depicted in Fig.~\ref{fig:fig2}(f). The FS exhibits a pronounced sublattice polarization, with dominant contributions from the $B$-sublattice sites ($5$ and $6$), whereas all $A$-sublattice sites ($1$–$4$) contribute relatively weaker, uniform, unpolarized weight. Here, it is worth noting that the aforementioned analysis assumes zero detuning, i.e., $\mu^{}_A=\mu^{}_B=0$. A positive (negative) detuning applied to the $4f$ sites decreases (increases) their sublattice contribution, and the corresponding weight vanishes in the limit $\mu^{}_A \rightarrow \infty$.

A qualitatively different type of sublattice polarization emerges when $t^{\prime}=t^{\prime\prime}=0.5$. The corresponding band structure, shown in Fig.~\ref{fig:fig2}(e), contains three VHSs. For a filling $n_{f}\approx0.68$, corresponding to the middle VHS (Fermi level shown by red line), the FS forms a square contour oriented along the BZ diagonals [see Fig.~\ref{fig:fig2}(c)]. The sublattice decomposition of the eigenstates, plotted in Fig.~\ref{fig:fig2}(g), reveals that this FS is dominated exclusively by contributions from the $A$-sublattice sites ($1$–$4$), while the $B$-sublattice sites are effectively absent. Moreover, the $A$-site contributions alternate along the sides of the square Fermi contour, producing an alternating pattern of sublattice polarization. Note that, unlike the previous scenario, the sublattice contributions are independent of the detuning parameters in this case. 

In summary, two distinct regimes of sublattice polarization emerge in the square–kagome metal: one dominated by the $B$ sublattice and another characterized by alternating $A$-sublattice contributions. In the following section, we demonstrate how these two types of sublattice polarization drive the formation of distinct AM orders, as discussed in Sec.~\ref{sec:symmetry_AM}.
\begin{figure}\includegraphics[width=1.0\linewidth]{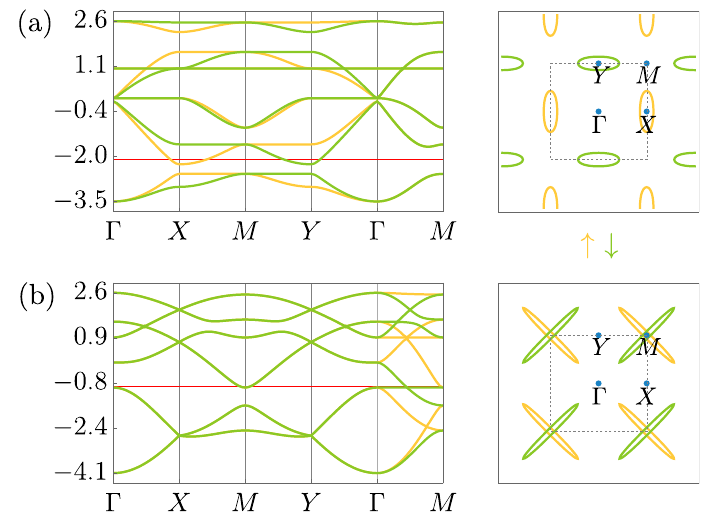}
\caption{Non-relativistic lifting of Kramers degeneracy for $U=5$: (a) $(t,t^\prime,t^{\prime\prime})=(1,0.5,0)$ and (b) $(t,t^\prime,t^{\prime\prime})=(1,0.5,0.5)$. The dashed square indicates the boundary of the first Brillouin zone.}
\label{fig:fig3}
\end{figure}
\section{AM orders}
\label{sec:am_orders}
Our analysis of the parent square-kagome metallic state in the preceding section provides an itinerant framework for exploring symmetry-broken magnetic phases. To realize the AM orders emerging from this itinerant background, we now consider magnetic instabilities induced by the on-site Hubbard interaction introduced in Eq.~\eqref{eq:model_int}. Within the mean-field approximation, the quartic term can be decoupled as  
\begin{equation}\label{eq:mf_decoupling_1}
    \hat{n}_{i,s,\uparrow} \hat{n}_{i,s,\downarrow}
    \approx
    (\hat{n}_{i,s,\downarrow} - \hat{n}_{i,s,\uparrow}) M_{i,s} + M^2_{i,s},
\end{equation}
where the local spin-polarization mean field $M_{i,s}$ is defined via  
\begin{equation}\label{eq:mf_decoupling_2}
    \langle \hat{n}_{i,s,\uparrow} - \hat{n}_{i,s,\downarrow} \rangle = 2M_{i,s}.
\end{equation}
Depending on which sublattice (A or B) acquires a finite $M_{i,s}$, distinct AM phases are stabilized. The corresponding saddle-point configurations of $M_{i,s}$ are obtained self-consistently at finite $U$. The results are summarized below.

\subsection{Sublattice-polarized eigenstates on $2c$ sites}
In the first case---corresponding to the sublattice-polarized eigenstates discussed in the previous section, where the dominant electronic weight resides on the B sites ($5$ and $6$)---the saddle-point solution yields
\begin{equation}
M_{i,1}=M_{i,2}=M_{i,3}=M_{i,4}\approx0,\quad
M_{i,5}=-M_{i,6}\neq0.    
\end{equation}
This solution indicates a magnetic instability in which only the B sites develop finite spin polarization with AFM order as shown in Fig.~\ref{fig:fig1}(b), while the A sites remain essentially nonmagnetic, despite their finite contribution to the electronic eigenstates.  
The resulting spin texture corresponds to a $d_{x^2-y^2}$-type AM order, as discussed in Sec.~\ref{sec:symmetry_AM}. Consequently, the spin degeneracy of the electronic bands [see Fig.~\ref{fig:fig2}(d)] is lifted, producing the non-relativistic spin-split band structure shown in Fig.~\ref{fig:fig3}(a). Notably, a residual degeneracy persists along the Brillouin-zone (BZ) diagonals ($\overline{\Gamma M}$), consistent with the symmetry analysis presented in Sec.~\ref{sec:symmetry_AM}.  

\subsection{Sublattice-polarized eigenstates on $4f$ sites}
In the second case---where the low-energy electronic eigenstates are polarized on the A sites ($1$--$4$) with negligible contribution from the B sites---the saddle-point configuration reads
\begin{equation}
M_{i,1}=-M_{i,2}=M_{i,3}=-M_{i,4}\neq0,\quad
M_{i,5}=M_{i,6}\approx0.    
\end{equation}
Here, an alternating spin polarization emerges exclusively on the A sublattice, forming an antiferromagnetic (AFM) pattern [see Fig.~\ref{fig:fig1}(c)]. This configuration realizes a $d_{xy}$-type AM order, as characterized in Sec.~\ref{sec:symmetry_AM}.  
Compared to the spin-degenerate metallic band structure shown in Fig.~\ref{fig:fig2}(e), the AM phase exhibits a spin-split spectrum displayed in Fig.~\ref{fig:fig3}(b). The remaining degeneracies along the BZ boundaries and axes ($\overline{\Gamma X}$, $\overline{\Gamma Y}$, $\overline{M X}$, and $\overline{M Y}$) align with the symmetry constraints derived in Sec.~\ref{sec:symmetry_AM}.  

In summary, a sublattice polarization on the A sites gives rise to a $d_{xy}$-type AM order, whereas polarization on the B sites stabilizes a $d_{x^2-y^2}$-type AM order. To further elucidate their low-energy properties, we derive effective models for the two AM phases by integrating out the nonmagnetic sites in each case. The effective model of AM for spin-polarization on the $2c$ Wyckoff positions (i.e., B sites) takes the form of Eq.~\eqref{eq:eff_BB} with $t_B=\frac{t^2}{2t'-\mu^{}_A}$, $t'_B=\frac{2t^2t'}{\mu^{}_A(2t'-\mu^{}_A)}$ and $\Tilde{\mu}^{}_B=\mu^{}_B-\frac{4t^2}{\mu^{}_A}$.
Similarly, the low-energy effective Hamiltonian for the other AM phase, after integrating the nonmagnetic sites (B sites) out, can be cast as Eq.~\eqref{eq:eff_AA} with $t^{}_A=-\frac{t^2}{{\mu}^{}_B}-t',\,t^{\prime}_A=t^{\prime\prime}_A=-\frac{t^2}{{\mu}^{}_B},\,\tilde{\mu}^{}_A=\mu^{}_A-\frac{2t^2}{{\mu}^{}_B}$.
The detailed derivations are provided in Appendix~\ref{app:effective}. Furthermore, we find that the $d_{x^{2}-y^{2}}$-type AM phase appears robustly for any choice of $t'$ when $t'' = 0$, indicating that this order is a stable phase of the system. Moreover, the $d_{xy}$-type AM order persists even for small deviations of $t''$ from $t'$. Although such deviations slightly distort the square Fermi surface, the altermagnetic order remains stable within our mean-field analysis.

\section{Stability Analysis}
\label{sec:stability_analysis}

To perform a stability analysis of the two AM phases, we employ the SRIKR slave boson formalism~\cite{{Kotliar-1986},{Fresard-1992}}. In this representation, the physical electron operator is mapped onto pseudofermions coupled to the auxiliary bosons via the SRIKR $z$-matrix
\begin{equation}
c^\dagger_{i\sigma} = \sum_{\sigma'} z^\dagger_{i,\sigma\sigma'}\, f^\dagger_{i\sigma'} \, .
\end{equation}
Here $f^\dagger_{i\sigma}$ creates a coherent quasiparticle, while $z^\dagger_{i,\sigma\sigma'}$ is the spin-rotationally invariant quasiparticle-weight matrix encoding the local Hubbard physics through the fields $e_i$, $d_i$, and $p_{i,\mu}$ with $\mu\in\{0,1,2,3\}$, describing empty, doubly  and singly occupied local configurations.

Since this mapping enlarges the Hilbert space, the original physical subspace has to be recovered by imposing local constraints, enforced in the path-integral formulation via the Lagrange multipliers $\alpha_i$, $\beta_{0,i}$, and $\beta_{i,a}$ with $a\in\{1,2,3\}$. The main benefit of our method is that by applying a meanfield approximation the Hubbard interaction becomes quadratic, at the cost of a more complex hopping term, which is no longer purely fermionic. Within this approximation the ground state is given by the saddle point of the free energy~\cite{Riegler-2020}. To describe the altermagnetic states considered in this work, we extend the standard single-site slave boson mean-field formalism to a cluster mean-field approach. In this scheme, the lattice is decomposed into clusters containing multiple, potentially inequivalent sites. Each site within the cluster is associated with its own set of slave boson fields, allowing for sublattice-dependent quasiparticle renormalization and magnetic polarization~\cite{clustersb}.

To calculate response functions, we consider Gaussian fluctuations around our saddle point by expanding the path-integral action $S$ up to second order in bosonic fields~\cite{Seufert-2021}
\begin{equation}
        \delta S^{(2)} = \sum_{q,n} \delta\psi_{s,\mu}(-\mathbf{q},-i\omega_n)\, \mathcal{M}^{s s'}_{\mu\nu}(\mathbf{q}, i\omega_n)\, \delta\psi_{s',\nu}(\mathbf{q},i\omega_n),
\end{equation}
with the fluctuation matrix $\mathcal{M}$

\begin{equation}
    \mathcal{M}_{\mu\nu}^{s s'}(\mathbf{q},i\omega_n) = \frac{\delta^2 S}{\delta\psi_{s,\mu}(-\mathbf{q},-i\omega_n)\;
                     \delta\psi_{s',\nu}(\mathbf{q},i\omega_n)} .
\end{equation}
 
The fluctuation fields within the sublattice basis can be straightforwardly related to the basis of magnetic fluctuations ($\delta \psi_{\bm q+ a \bm Q}$) used in the previous publication~\cite{Seufert-2021} by Fourier-transform and an additional phase factor
{
    \begin{equation}
        \delta \psi_{\bm q+ a \bm Q} = \sum_\alpha \delta \psi_\alpha e^{i (\bm q + a \bm Q) \bm r_\alpha}.
    \end{equation}
}

However, there are some differences such as that the current basis choice also encompasses the charge SB degrees of freedom in the numerical treatment of the resulting loop expansion. This goes beyond the scope of the current work and will be published in full detail in a follow-up paper~\cite{clustersb}. 

From the Gaussian fluctuation action, linear response functions are obtained by evaluating correlation functions of the fluctuating bosonic fields. In frequency and momentum space, the generalized susceptibility is given by the inverse of the fluctuation matrix,

\begin{align}
    \chi^{s s'}_{\mu\nu}(q,i\omega_n) &= \left\langle \delta\psi_{s,\mu}(-\mathbf{q},-i\omega_n)\,
                                          \delta\psi_{s',\nu}(\mathbf{q},i\omega_n) \right\rangle \nonumber  \\
                                      &= \left[ \mathcal{M}^{-1}(\mathbf{q},i\omega_n) \right]^{s s'}_{\mu\nu} .
\end{align}

Due to spin-rotation invariance, the fluctuation matrix $\mathcal{M}_{\mu\nu}$ is block diagonal in charge and spin sectors. The charge susceptibility is defined as the density--density response,
\begin{equation}
\chi^{ss'}_c(\mathbf{q},i\omega_n)
=
\langle
\delta n_s(-\mathbf{q},-i\omega_n)\,
\delta n_{s'}(\mathbf{q},i\omega_n)
\rangle,
\end{equation}
and is obtained from the charge block of $\mathcal{M}^{-1}$.

Similarly, the spin susceptibility is defined as the spin--spin correlation function,

\begin{equation}
    \chi^{s s'}_{\mathrm{s}}{}^{\alpha\beta}(\mathbf{q},i\omega_n)
=
\left\langle
\delta S^{\alpha}_{s}(-\mathbf{q},-i\omega_n)\,
\delta S^{\beta}_{s'}(\mathbf{q},i\omega_n)
\right\rangle ,
\end{equation}

where the spin density is expressed in terms of the slave boson fields as
\begin{equation}
\mathbf{S}_s = \tilde{\mathbf{p}}_s p_{0,s} ,
\qquad
\tilde{\mathbf{p}}_s = (p_{x,s},-p_{y,s},p_{z,s})^T .
\end{equation}

With the formalism set up, we can now calculate susceptibilities for each AM phase and determine their stability. 

\begin{figure}
    \centering
    \includegraphics[width=1.0\linewidth]{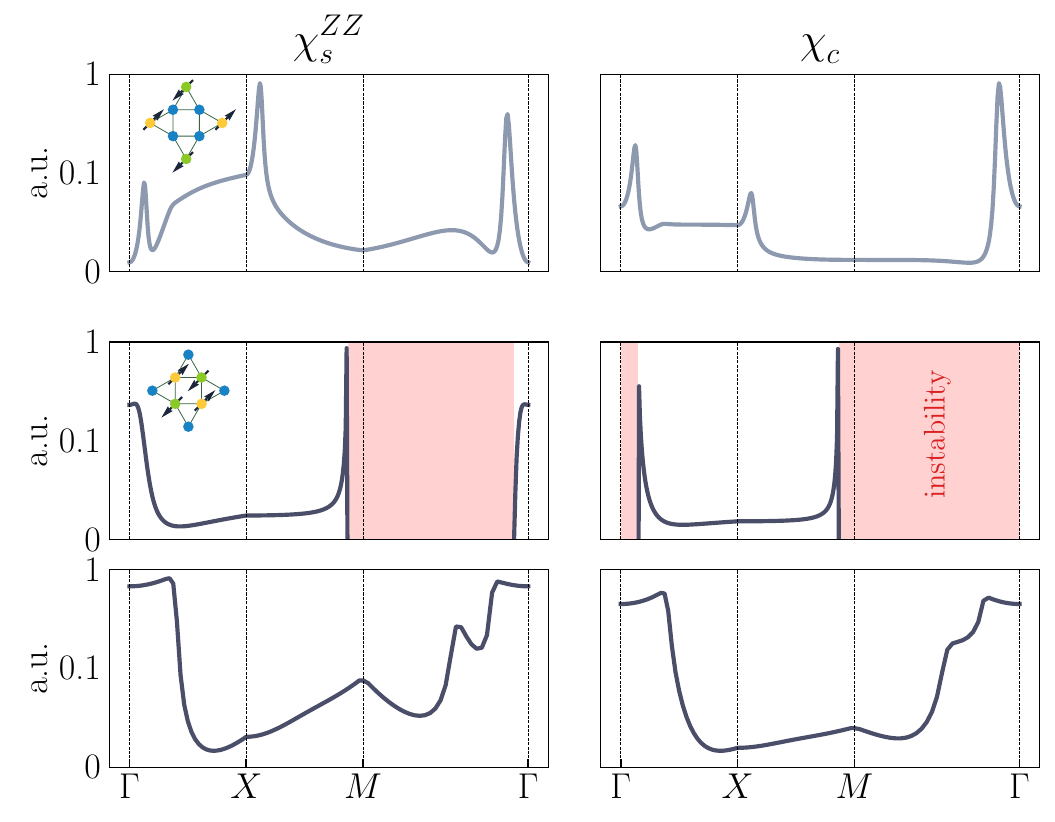}
    \caption{Stability analysis for both altermagnetic saddle points with momentum-resolved response functions in arbitrary units (a.u.) calculated within the SRIKR Gaussian-fluctuation analysis. Shown along the high-symmetry path $\Gamma$--X--M--$\Gamma$ are the longitudinal spin susceptibility $\chi^{zz}_s$ (left), and charge susceptibility $\chi_c$ (right). The rows correspond to: (top) AFM order on B sites (Wyckoff $2c$) (stable, no instabilities), (middle) AFM order on A sites (Wyckoff $4f$) at $V=0$ (divergent charge/longitudinal-spin response indicating instability), and (bottom) AFM order on A sites with nearest-neighbor density interaction $V=2$ (suppressed instabilities indicating stabilized homogeneous phase).}
    \label{fig:chi_plot}
\end{figure}

In Fig.~\ref{fig:chi_plot}, the longitudinal spin susceptibility ($\chi_s^{zz}$), together with the charge susceptibility ($\chi_c$), are shown for both altermagnetic configurations along the high symmetry path. Since divergences in the corresponding response functions indicate an instability of the saddle-point solution towards charge- or spin-ordered states, the momentum-resolved structure of these susceptibilities allows us to assess the stability of the respective phases. In the following analysis, the transverse spin susceptibilities ($\chi_s^{xx}$, $\chi_s^{yy}$) are neglected, as no instabilities are observed for either state.

\textit{AFM order on B sites (Wyckoff $2c$):}
The susceptibilities corresponding to this configuration are displayed in the first row of Fig.~\ref{fig:chi_plot}. In the longitudinal spin channel $\chi_s^{zz}$ as well as in the charge channel $\chi_c$, several smaller peaks are visible at certain momenta. These features signal enhanced fluctuations and may indicate a tendency toward competing ordering phenomena. However, none of these resemble an instability, since the susceptibilities remain finite and positive over the entire Brillouin zone. We therefore conclude that, for the chosen parameter set this state is stable against both magnetic and charge fluctuations.

\textit{AFM order on A sites (Wyckoff $4f$):}
The stability analysis for this configuration is more subtle. The second row of Fig.~\ref{fig:chi_plot} show the susceptibilities at the filling $n = 0.68$. In contrast to the previous case, pronounced negative divergences appear in the longitudinal spin susceptibility $\chi_s^{zz}$ and, even more prominently, in the charge susceptibility $\chi_c$. The presence of these divergences indicate that the saddle-point solution is unstable with respect to additional ordering tendencies. In particular, the negative peak of $\chi_c$ at the $\Gamma$ point strongly suggests a tendency toward macroscopic phase separation~\cite{Seufert-2021}. To verify this, we analyzed the behavior of the chemical potential $\mu_0$ as a function of filling $n$. In the presence of phase separation, the electronic compressibility becomes negative, which is reflected in a negative slope $\partial \mu_0 / \partial n < 0$. Indeed, we find that the $\mu_0(n)$ curve exhibits a region with negative derivative [see Fig.~\ref{fig:maxwell} of Appendix.~\ref{maxwell_con}] which is consistent with the instability against phase separation indicated by the negative divergence in $\chi_c$ at $\Gamma$. In such a situation, the homogeneous solution is thermodynamically unstable and a Maxwell construction is required to determine the coexistence region of the phase separated state. In the present case, this suggests phase separation between the state with AFM order on the A sites and a paramagnetic state. 

To further verify this interpretation, we included an additional nearest-neighbor density interaction $V$ between the A and B sites, which is known to counteract macroscopic phase separation by penalizing charge inhomogeneities~\cite{Riegler-2023}. As expected, this nonlocal interaction suppresses the instability. In particular, for $V = 2$ (see last row of Fig. \ref{fig:chi_plot}), all divergences in the charge and longitudinal spin susceptibility disappear entirely. Consistently, the region of negative compressibility is removed. This can be seen as additional evidence that the instability observed at $V=0$ is indeed associated with phase separation since sufficiently strong nearest-neighbor interactions stabilize the homogeneous phase.

Even for $V = 0$, the instability is only restricted to a narrow filling window as can be seen from the maxwell construction. 
The chemical potential $\mu_0(n)$ exhibits a region of negative slope only in the interval 
$n \approx 0.666\text{-}0.694$. Outside this range, the derivative 
$\partial \mu_0 / \partial n$ is positive again, indicating that 
the homogeneous phase regains thermodynamic stability. Consequently, 
the tendency toward phase separation disappears around $n \approx 0.694$. This is fully consistent with the behavior of the charge susceptibility where 
the divergence at $\Gamma$ is confined to the same filling regime 
and vanishes for larger $n$. Thus, for sufficiently high fillings, the 
homogeneous AM state remains stable even without a nearest-neighbor density interaction.

We also investigated the behavior of the system upon a slight variation of the hopping parameter to $t'' = -0.4$. 
In this case, the region of phase separation is not removed but shifted to larger fillings, now extending over the interval 
$n \approx 0.685\text{--}0.705$. This indicates that the tendency toward phase separation is robust with respect to moderate 
changes of $t''$, while its precise location in filling is sensitive to the band-structure details. As in the previously discussed case, the phase separation corresponds to coexistence between the AFM state with order on the A sites and a paramagnetic phase. Including a nearest-neighbor density interaction $V$ again suppresses this instability and restores the 
stability of the homogeneous solution.

\section{Discussion and Outlook}
\label{sec:discussion}
In this work, we investigate the emergence of AM phases in a two-sublattice itinerant system on the square–kagome lattice. Starting from two symmetry-allowed AM phases and their corresponding minimal models, we introduce a single-orbital Hubbard model and demonstrate that the interplay between lattice geometry and sublattice-polarized electronic eigenstates drives these AM instabilities even in the absence of explicit spin–orbit coupling or orbital ordering. Our results show that the AM phases originate from a purely kinetic mechanism: sublattice interference encoded in the Bloch wave functions produces an effective breaking of time-reversal symmetry in momentum space once magnetic order develops on a subset of sublattices.

By analyzing the noninteracting band structure, we identified two regimes of sublattice polarization corresponding to different van Hove fillings. 
In one case, the low-energy eigenstates are dominated by the $2c$-Wyckoff sites, leading to a $d_{x^{2}-y^{2}}$-type AM phase once magnetic instability sets in. 
In the complementary regime, the $4f$-Wyckoff sites dominate the Fermi surface, giving rise to a $d_{xy}$-type AM order characterized by alternating sublattice magnetizations. 
Both phases exhibit nonrelativistic spin splitting in the electronic spectrum while preserving zero net magnetization, in accordance with the symmetry classification discussed earlier. 
Importantly, our mean-field results establish that the $d_{x^{2}-y^{2}}$ AM phase remains stable across a broad parameter range, while the $d_{xy}$ AM order persists for small deviations between $t'$ and $t''$, demonstrating the robustness of these symmetry-distinct altermagnetic states.

Beyond the mean-field level, the stability of the AM phases is governed by their response to collective charge and spin fluctuations. Our Gaussian fluctuation analysis within the SB framework shows that the $d_{x^2-y^2}$-type AM phase, driven by magnetic order on the Wyckoff-$2c$ sites, is stable, as all charge and spin susceptibilities remain finite throughout the Brillouin zone. By contrast, the $d_{xy}$-type AM phase associated with order on the Wyckoff-$4f$ sites exhibits an instability in the charge and longitudinal spin channels, indicating a tendency toward phase separation in a narrow filling regime. This instability can  either be suppressed by moderate nonlocal interactions, restoring the stability of the homogeneous altermagnetic state or vanishes for larger filling $n$.

The present study provides a concrete microscopic realization of the sublattice-interference–driven AM mechanism recently proposed in the context of the Lieb lattice~\cite{Matteo-2025}. 
Our findings thus extend this paradigm to a new lattice geometry and confirm that sublattice polarization of itinerant electrons can act as a universal route to altermagnetism in multiband systems. 
Moreover, the coexistence and competition between multiple AM symmetry channels, as found here, suggest the possibility of AM phase transitions tuned by band structure parameters, electron filling, or external perturbations such as strain and pressure.

Looking forward, several open directions naturally emerge from this work. Beyond the mean-field level, it will be important to study fluctuation effects and assess the possible coexistence or competition with other electronic instabilities, such as density-wave or unconventional magnetic phases. Furthermore we can extend the present SB analysis to the calculation of dynamical susceptibilities, which would give access to the collective spin and charge excitation spectrum beyond the static Gaussian fluctuations. In particular, this would allow us to study the magnon bands of the AM states and to identify characteristic features tied to the symmetry-distinct $d_{x^2-y^2}$ and $d_{xy}$ orders~\cite{{Sinova-2024},{Smejkal-2023}}. The topological aspects of the spin-split altermagnetic bands~\cite{Rafael-2025,Roig-2024}, including Berry curvature and anomalous transport signatures, also remain to be explored. Although the material realization of the square-kagome lattice has been achieved primarily in the localized-spin (Mott) regime~\cite{Fujihala-2020,gonzalez-2025,Niggemann-2023}, explicit realizations in an itinerant-electron setting remain comparatively scarce. Nevertheless, alternative routes may provide viable platforms. In particular, metal–organic frameworks~\cite{Huang-2024} and cold-atom optical lattices~\cite{Das-2024} offer highly tunable lattice geometries and hopping amplitudes, potentially enabling experimental access to both the $d_{x^{2}-y^{2}}$ and $d_{xy}$ AM regimes. Overall, our results establish the square–kagome lattice as a fertile ground for realizing and tuning symmetry-distinct AM phases, thereby broadening the scope of itinerant AM and paving the way toward engineered spin-polarized states without net magnetization.

\begin{acknowledgements}
We thank Hendrik Hohmann, Matteo D\"urrnagel and Lennart Klebl for useful discussions. This work is supported by the Deutsche Forschungsgemeinschaft (DFG, German Research Foundation) through Project-ID 258499086 -- SFB 1170 and through the W\"urzburg-Dresden Cluster of Excellence on Complexity, Topology and Dynamics in Quantum Matter -- ctd.qmat, Project-ID 390858490 -- EXC 2147.
\end{acknowledgements}
\appendix
\section{EFFECTIVE MODELS}
\label{app:effective}
In this Appendix, we derive a low-energy effective Hamiltonian for the
magnetically ordered sublattice $s$ by integrating out the degrees of
freedom associated with the nonmagnetic sublattice $s'$.
 Starting from the tight-binding Hamiltonian in sublattice space,
we write
\begin{equation}
    H =
    \begin{pmatrix}
        H_{ss} & H_{ss'} \\
        H_{ss'}^\dagger      & H_{s's'}
    \end{pmatrix} ,
\end{equation}
where $H_{ss}$ and $H_{s's'}$ act within the $s$ and $s'$ subspaces,
respectively, and $H_{ss'}$ describes the hybridization between them.
The corresponding free Green's functions are
$\mathcal{G}_{ss}^0 = (i\omega - H_{ss})^{-1}$ and
$\mathcal{G}_{s's'}^0 = (i\omega - H_{s's'})^{-1}$.
We now construct an effective description for the $s$ sublattices by
integrating out the $s'$ sector.
Using standard projection / decimation techniques
\cite{karrasch2010functional,Kalkstein1971,LopezSancho1985},
the $s$-subspace Green's function can be written as a continued fraction
\begin{equation}
\begin{split}
    \mathcal{G}_{ss}
   &= \frac{1}{
    \big(\mathcal{G}_{ss}^0\big)^{-1} 
    - H_{ss'} \frac{1_{\phantom{g}}}{\big(\mathcal{G}_{s's'}^0\big)^{-1} - H_{s'\cdot} \cdots \, H_{\cdot \, s'}^\dagger} H_{ss'}^\dagger
    } \,.
\end{split}
\end{equation}
In practice, we approximate this by resumming the leading local process,
which yields an effective Green's function
\begin{align}
    \mathcal{G}_{ss}(i\omega)
    \approx&
    \frac{1}{i\omega - H_{ss} - H_{ss'}\,\mathcal{G}_{s's'}^0(i\omega)\,H_{ss'}^\dagger}\notag\\
    &\equiv
    \frac{1}{i\omega - H_{ss} - \Sigma_{ss}(i\omega)} \, .
\end{align}
where $\Sigma_{ss}(i\omega)=H_{ss'}\,\mathcal{G}_{s's'}^0(i\omega)\,H_{ss'}^\dagger$ is the hybridization or self-energy. By calculating this, an effective Hamiltonian in the zero frequency limit, reads as:
\begin{equation}
H^{\text{eff}}_s = H_{ss}
+\Sigma_{ss}(i\omega\rightarrow0),
\label{eq:Heff_SW}
\end{equation}

In square-kagome lattice there are two sublattice, i.e., $s,s'=\{A,B\}$ and the Hamiltonian reads as 
\begin{equation}
H=
    \begin{bmatrix}
H^{}_{AA} & H^{}_{AB} \\
H^{\dagger}_{AB} & H^{}_{BB}
\end{bmatrix}
\end{equation}
where
\begin{equation}
H_{AA}=
\begin{bmatrix}
\mu_A & -t' e^{-i a k_y}  & 0 & -t' e^{i a k_x} \\
-t' e^{i a k_y} & \mu_A & -t' e^{i a k_x}  & 0  \\
0 & -t' e^{-i a k_x} & \mu_A & -t' e^{i a k_y}  \\
-t' e^{-i a k_x} & 0 & -t' e^{-i a k_y} & \mu_A
\end{bmatrix}
,
\end{equation}
\begin{equation}
    H^{}_{BB}=\mu^{}_B \mathrm{I}_{2\times2}
\end{equation}
and 
\begin{equation}
H^{}_{AB}=-t
    \begin{bmatrix}
 e^{-i \frac{1}{2} ( k^{}_xb + k^{}_ya)} & e^{i \frac{1}{2} (k^{}_xa + k^{}_y b)} \\
 e^{-i \frac{1}{2} (k^{}_xb - k^{}_ya)} & e^{i \frac{1}{2} (k^{}_xa - k^{}_yb)} \\
 e^{i \frac{1}{2} (k^{}_xb + k^{}_ya)} & e^{-i \frac{1}{2} (k^{}_xa + k^{}_yb)} \\
 e^{i \frac{1}{2} ( k^{}_xb - k^{}_ya)} & e^{-i \frac{1}{2} (k^{}_xa - k^{}_yb)}
\end{bmatrix}
\,.
\end{equation}
In the following we calculate the low energy effective models for the two AM phases.
\subsection{ FOR THE A SITES}
In this case $s=A$ and $s'=B$. Eq.~\eqref{eq:Heff_SW} gives the effective model for $d_{xy}$ AM phase.
\begin{equation}
H^{\text{eff}}_{\text{A}}=
    \begin{bmatrix}
 H^{d}_{\text{A}} & H^{o}_{\text{A}} \\
 (H^{o}_{\text{A}})^\dagger & (H^{d}_{\text{A}})^T
\end{bmatrix}
,
\end{equation}
where
\begin{widetext}
\begin{equation}
H^{d}_{\text{A}}=
\begin{bmatrix}
\tilde{\mu}^{}_A & t^{}_A e^{-i a k^{}_y} + t'_A e^{i b k^{}_y} \\
t^{}_A e^{i a k^{}_y} + t'_A e^{-i b k^{}_y} & \tilde{\mu}_A 
\end{bmatrix}
,\,t^{}_A=-\frac{t^2}{{\mu}^{}_B}-t',\,t^{\prime}_A=-\frac{t^2}{{\mu}^{}_B},\,\tilde{\mu}^{}_A=\mu^{}_A-\frac{2t^2}{{\mu}^{}_B},
\end{equation}
 and 
\begin{equation}
 H^{o}_{\text{A}}=
\begin{bmatrix}
t''_A \left(e^{-i  (k^{}_xb + k^{}_ya)} + e^{i  (k^{}_xa + k^{}_yb)}\right) & 
t^{}_A e^{i a k^{}_x} + t'_A e^{-i b k^{}_x} \\
t^{}_A e^{i a k^{}_x} + t'_A e^{-i b k^{}_x} & 
t''_A \left(e^{-i  (k^{}_xb - k^{}_ya)} + e^{i  (k^{}_xa - k^{}_yb)}\right)
\end{bmatrix}
\,,\,t''_A=-\frac{t^2}{{\mu}^{}_B}\, .
\end{equation}
\end{widetext}
\subsection{ FOR THE B SITES}
In this case $s=B$ and $s'=A$. Eq.~\eqref{eq:Heff_SW} give the effective model for $d_{x^2-y^2}$ AM phase.
\begin{equation}
{H^{\text{eff}}_{\text{B}}} = 
\begin{bmatrix}
\Tilde{\mu}^{}_B + t'_{B} \cos(k^{}_x\Tilde{a}) & 
t_B \cos\left(\frac{k^{}_x\Tilde{a}}{2}\right) \cos\left(\frac{k^{}_y\Tilde{a}}{2}\right) \\
t_B \cos\left(\frac{k^{}_x\Tilde{a}}{2}\right) \cos\left(\frac{k^{}_y\Tilde{a}}{2}\right) & 
\Tilde{\mu}^{}_B + t'_{B} \cos(k^{}_y\Tilde{a})
\end{bmatrix}
\end{equation}
where $\Tilde{a}=a+b$, $t_B=\frac{t^2}{2t'-\mu^{}_A}$, $t'_B=\frac{2t^2t'}{\mu^{}_A(2t'-\mu^{}_A)}$ and $\Tilde{\mu}^{}_B=\mu^{}_B-\frac{4t^2}{\mu^{}_A}$.

\begin{figure}[h!]
    \centering
    \includegraphics[width=0.8\linewidth]{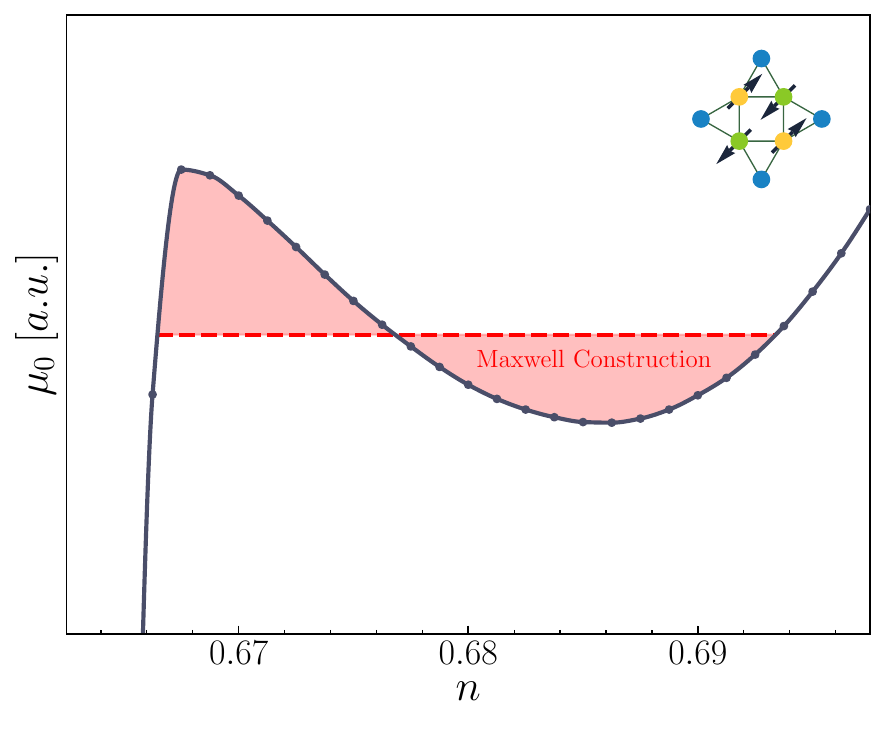}
    \caption{Chemical potential $\mu_0$ versus filling $n$ in the unstable regime of the A-site (Wyckoff $4f$) altermagnetic phase at $V=0$. The region with negative slope $\partial \mu_0/\partial n < 0$ signals negative compressibility and a tendency toward macroscopic phase separation. The dashed line indicates the Maxwell construction used to determine the coexistence window.}
    \label{fig:maxwell}
\end{figure}

\section{MAXWELL CONSTRUCTION}
\label{maxwell_con}
To determine the filling interval in which a homogeneous solution becomes thermodynamically unstable, we perform a Maxwell construction on the chemical–potential curve $\mu_{0}(n)$. In the regime where the
mean-field solution exhibits negative compressibility, $\partial \mu_{0} / \partial n < 0$, the
system lowers its free energy by separating into two coexisting phases with fillings
$n_{1}$ and $n_{2}$. These coexistence points are obtained by replacing the non-monotonic
segment of $\mu_{0}(n)$ with a horizontal line such that the enclosed areas above and
below this line are equal. The Maxwell construction shown in Fig.~\ref{fig:maxwell} thereby identifies the boundaries
of the phase-separated region and determines the density range in which the
homogeneous state is not energetically stable.

\bibliography{references}
\end{document}